# Experimental Tests of the Proportionality of Aerodynamic Drag to Air Density for Supersonic Projectiles

Elya Courtney, Amy Courtney, Michael Courtney
BTG Research, 9574 Simon Lebleu Road, Lake Charles, LA, 70607
michael_courtney@alum.mit.edu

**Abstract**

Pure theory recognizes the dependence of supersonic drag coefficients on both Mach number and Reynolds number, which includes an implicit dependence of drag coefficient on air density. However, many modern approaches to computing trajectories for artillery and small arms treat drag coefficients as a function of Mach number and assume no dependence on Reynolds number. If drag force is strictly proportional to air density for supersonic projectiles (as suggested by applied theory), the drag coefficient should be independent of air density over a range of Mach numbers. Experimental data to directly support this are not widely available for supersonic projectiles. The experiment determined drag on a 2.59 g projectile from M1.2 to M2.9 using optical chronographs to measure initial and final velocities over a separation of 91.44 m. The free flight determination of drag coefficients was performed at two significantly different atmospheric densities (0.93 kg/m$^3$ and 1.15 kg/m$^3$). This experiment supported direct proportionality of aerodynamic drag to air density from M1.2 to M2.9 within the experimental error of 1%-2%.

## 1. Introduction

Both subsonic and supersonic projectiles experience drag, that is, air resistance, that causes them to slow down with time [1]. Knowing how drag affects an object in flight is essential for predicting the flight path [2]. The assumption that drag force is proportional to air density is also used to infer the air density from other measurements [3]. The theoretical relationship (equation) relating the drag force and different physical and environmental characteristics is widely available:

$$F_d = \frac{1}{2} C_d A V^2 \rho \quad , (1)$$

where $C_d$ is the drag coefficient, $A$ is the cross-sectional area, $V$ is the flow velocity, $\rho$ is air density, and $F_d$ is the drag force [4].

This equation suggests that drag force is affected by air density and specifically that drag force is directly proportional to air density. This proportionality is only disturbed if the drag coefficient depends on air density explicitly, or implicitly through the Reynolds number. In general, drag coefficients for supersonic projectiles depend on both the Mach number and the Reynolds number [1], and both these factors are routinely considered in many applications such as using balloon accelerations to determine atmospheric density and performing calculations of vehicles re-entering the earth's atmosphere.

Calculations of trajectories for artillery and small arms projectiles has been dominated by six degree of freedom numerical models and modified point mass models [2,5-8]. Possible dependence of drag coefficients on air density was retained in early models [9]. As theory was simplified for computational expedience [10], it was recognized that skin friction drag and base drag depend on Reynolds number as well as Mach number.

Skin friction drag only accounts for 8% to 16% of the total drag coefficient. McCoy [10,11] built in Reynolds number dependence into his codes for computing skin friction drag for both turbulent and laminar boundary layers, but his approach was to estimate Reynolds number as a constant times the Mach number times the projectile total length. This Reynolds number estimate for sea level conditions and a temperature of 15°C only accounted for Reynolds number variations for changes in Mach number and bullet length (See Eqn. 17 of McCoy [10]). Changes in fluid density or viscosity were not accounted for. Thus, this widely adopted approach to estimating skin friction drag, while in principle was dependent on Reynolds number, had implicitly eliminated possible dependence on air density, and when added to other components to estimate the total drag coefficient, had the result of keeping theoretical drag coefficients independent of air density.

Base drag can account for as much as 35% of the total drag coefficient, and can depend on Reynolds number (thus air density) in principle. McCoy's approach [10] to estimating base drag was to subtract all the other theoretical components from the measured total drag coefficient. He found no correlation between resulting experimental estimates for base drag and Reynolds number of a "large amount of high quality free-flight data" from projectiles tested by the Ballistics Research Laboratory. Consequently, this widely adopted computational approach for estimating base drag was independent of Reynolds number.

McCoy's computational approach yields theoretical drag coefficients that are independent of air density. Many approaches to using measured drag coefficients to predict trajectories using six degree of freedom and modified point mass models also neglect possible dependence of drag coefficients on air density. These theoretical models and computational approach have been tested against a large volume of data generated at the Ballistic Research Laboratory located at Aberdeen Proving Ground. However, since these experimental tests were conducted at Aberdeen, Maryland, at an approximate elevation of 29m above sea level, they do not represent independent validation of models that assume that drag coefficients are independent of air density. This paper presents an experimental test of this broadly applied framework that employs supersonic drag coefficients that are independent of air density. Experimental drag coefficients are usually calculated when the other factors in Equation 1 are measured experimentally. To test the theoretical relationship between drag force and air density at supersonic speeds, a 2.59 g projectile was launched at six different velocities between Mach 1.2 and Mach 2.9 and at two air densities.

## 2. Materials and Methods

The experimental design to determine drag coefficients used two CED Millennium optical chronographs with LED sky screens. After calibration, the chronographs were placed 3.04 meters and 94.48 meters from the muzzle. Chronograph separations were measured with a tape measure to within a few centimeters. Calibrations were performed by placing the two chronographs in a row, with minimal separation, and shooting though them. Each reading of the second chronograph is adjusted appropriately for the small loss of velocity (< 2 m/s) over the 0.6 m distance from the closest chronograph. Then the average velocity of ten shots was compared to determine systematic variations in the readings between the chronographs. This way, the variations between chronographs were reduced to 0.1%. The calibration procedure was repeated on each day of the experiment. One of the optical chronographs is shown in Fig. 1.

The equation used to experimentally determine the drag force from velocities measured over a distance, $d$, is based on the work-energy theorem and is

$$F_d = \frac{1}{2}\frac{m(v_f^2 - v_i^2)}{d}, \quad (2)$$

where $F_d$ is the drag force, $m$ is the projectile mass, $d$ is the distance between chronographs, and $v_f$ and $v_i$ are the final and initial velocities over the interval, respectively.

The equation that was used to compute the drag coefficient, $C_d$, is

$$F_d = \frac{1}{2}\rho v^2 C_d A, \quad (3)$$

where $F_d$ is the drag force, $\rho$ is the air density, $v$ is the average velocity over the interval (which has been substituted for the fluid flow velocity relative to the surface), and $A$ is the cross sectional area. Combining equations 1 and 2 to compute the drag coefficient from the measured velocities gives

$$C_d = \frac{2F_d}{Av^2\rho} = \frac{m(v_f^2 - v_i^2)}{dAv^2\rho}. \quad (4)$$

The drag coefficients at different Mach numbers were used to generate a graph of drag coefficient vs. Mach number.

Data was taken at two locations. The altitudes of the two locations were 0 meters and 1981 meters above sea level. Significantly different altitudes allowed testing at significantly different air densities. Environmental data (temperature, air pressure, and humidity) were measured with a Kestrel 4500 weather meter. The air densities were computed using the JBM Ballistics Calculator [12], which uses the temperature, pressure, and humidity at each location.

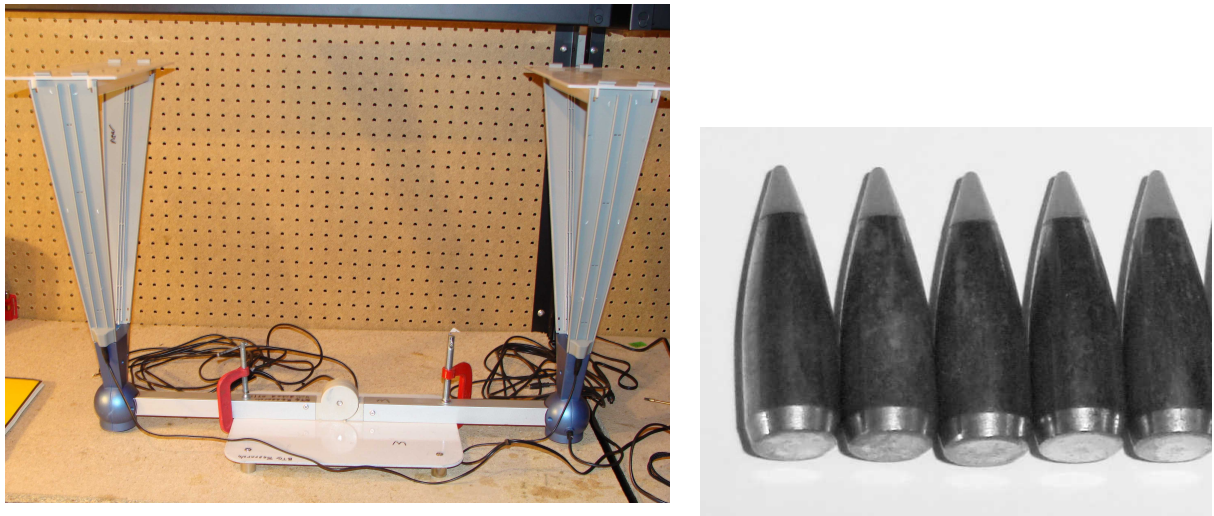

Fig. 1. One of the CED Millenium chronographs with LED skyscreens is shown (left) along with a sample of the Nosler Ballistic Tip bullets (right) used in the experiment.

To achieve a range of muzzle velocities from Mach 1.2 to Mach 2.9, 2.59 gram Nosler Ballistic Tip bullets (Part # 39510, Nosler, Inc., Bend, Oregon) were loaded in .223 Rem Lapua brass in front of 6, 8, 10, 11, 12, and 14 grains of Alliant Blue Dot powder and 29 grains of Hodgdon CFE 223 powder.

Ten shots each were fired for each powder charge except for 6 grains, for which 20 shots were fired. The rifle used was a Remington 700 ADL with a 1 in 12 inch twist. This bullet was chosen for testing because earlier testing (unpublished data) had shown it to have excellent shot to shot consistency in drag and also to have good accuracy to reduce the chances of accidentally shooting the far chronograph. More shots (20) were used at the lowest powder charge, because it was believed that the transonic drag rise might introduce more shot to shot variations in the drag coefficient. A picture of some of the bullets is shown in Fig. 1.

Velocity and air density were then used to determine drag coefficients over the interval for each shot. The mean and uncertainty (standard error of the mean) of the drag coefficient for each powder charge was then determined using standard statistical methods.

## 3. Results

Table 1 shows the environmental conditions at the two experimental locations. Temperature, pressure, and humidity were measured directly. Air density and speed of sound were computed from measured atmospheric conditions. The uncertainty in air density was determined by error propagation of the uncertainties in temperature, pressure, and humidity.

Table 1
Environmental conditions for each experimental location and projectile.

| Condition | Temperature (° C) | Humidity (%) | Pressure (mm Hg) | Density ($kg/m^3$) | Uncertainty in Density ($kg/m^3$) | Speed of Sound (m/s) |
|---|---|---|---|---|---|---|
| Low Air Density | 26.7 | 22.6 | 602.2 | 0.9293 | 0.0016 | 347.1 |
| High Air Density | 29.4 | 40.0 | 755.7 | 1.1526 | 0.0003 | 348.7 |

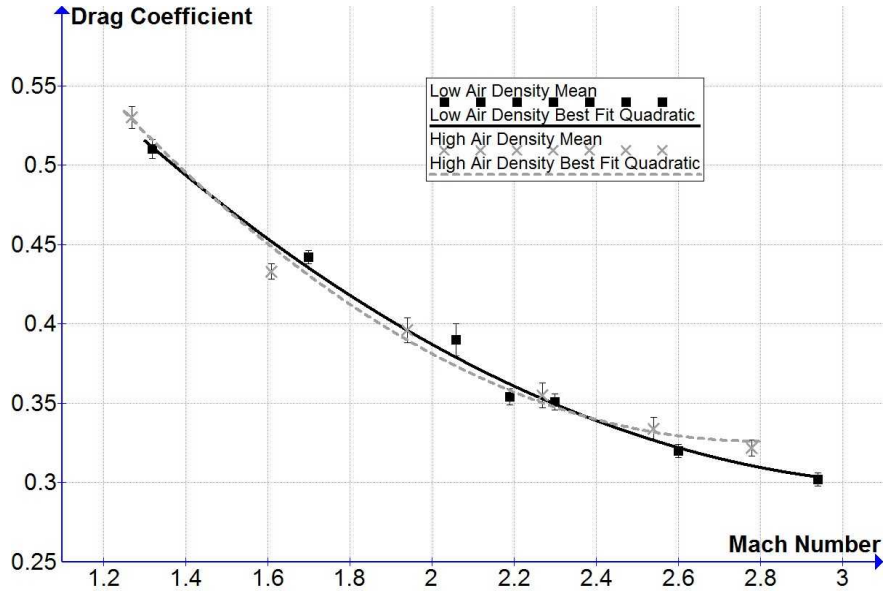

Fig. 2. Drag coefficient vs. Mach number at each air density shown with the best quadratic fit to the raw data (all shots for each data set). High Air Density: $R^2= 0.9521$. Low Air Density: $R^2=0.9618$. Tests were performed at a low air density of 0.9293 kg/m$^3$ and a high air density of 1.1526 kg/m$^3$.

Fig. 2 shows the results of ballistic testing and computation of experimental drag coefficients with increasing Mach number. As expected, the drag coefficient decreased as Mach number increased [10]. As Mach number increased from approximately 1.2 to 2.9, drag coefficient decreased from about 0.54 to 0.33. Within the small uncertainties (1% to 2%), results were similar for tests performed at higher and lower air densities. The gap between the fit lines for the data may seem like disagreement; however, the uncertainties of the curves overlap each other, so that no significant disagreement was observed.

## 4. Discussion and Conclusions

The original hypothesis, suggested by McCoy [10,11] and widely implemented in trajectory models, was that the drag force on a supersonic or subsonic projectile is proportional to air density. This suggestion depends on the drag coefficient being independent of air density, though it is known to depend on Mach number. The results of the supersonic experiment showed that the drag coefficient decreased as Mach number increased from about Mach 1.2 to Mach 2.9. However, results of tests performed at different air densities were not significantly different, smaller than the uncertainties of 1%-2%. Therefore the hypothesis that drag force on a supersonic projectile is proportional to air density was supported within the experimental uncertainties.

The uncertainties in the mean values of each supersonic drag coefficient were under 2%. The small uncertainties are probably due to three factors of the experimental design: 1) the projectile used for the experiments was chosen because it had shown consistent performance between tests in previous experiments; 2) for each powder charge, ten (or twenty) samples were tested at each air density; and 3) equipment used for critical measurements (velocity and environmental conditions) were accurate and precise.

McCoy's (1981) [10] analysis of Ballistics Research Laboratory data suggested that only the skin friction depended on Reynolds number, as he found no correlation between base drag and Reynolds number. Various formulas [13] estimate skin friction as depending on Reynolds number to the negative 1/3 to negative 1/5 power. This would imply that a 25% change in Reynolds number (as estimated in this experiment) would only yield a 5% to 8% change in skin friction. If skin friction accounts for 10% or less of the total drag, then the effect of changing Reynolds number would be expected to be less than 1% of the measured drag coefficient. In contrast, base drag is often a much more significant component to total drag than skin friction. Strong dependencies of base drag on Reynolds number would have been apparent in the experimental data, especially in the data below Mach 1.5.

Prior to this study, little published data were available to directly support or refute the proportionality of drag force and air density for supersonic projectiles. The results provide experimental support for using the drag force models to predict the flight path for artillery and small arms projectiles. The experimental method of measuring free flight near and far velocities was shown to be useful for measuring drag coefficients for both supersonic and subsonic projectiles precisely under different conditions without requiring wind tunnels or elaborate test and measurement equipment.

Since the projectiles were rotating in flight, there may be concern that rotational motion could influence air drag. While the projectiles had gyroscopic stabilities above 1.4 for the whole range of Mach numbers, it might still be possible that skin friction drag was influenced by rotation rates that varied with muzzle velocity. However, since the speed of sound was nearly the same at both locations due to nearly the same temperatures, any small variations in skin friction drag due to variations in rotation rate would have tracked closely with Mach number and be unlikely to contribute to drag variations observed between the two air densities.

This project had both strengths and weaknesses. The strengths included accuracy due to projectile selection and chronograph calibration, as well as having good sample sizes. This experiment also covered a range of Mach numbers, so the results could be applicable to more than just one type of projectile. A limitation to this experiment was having relatively a small change in atmospheric densities compared to what is theoretically possible.

In summary, few published experimental data are available to directly support whether drag coefficient depends on air density for supersonic projectiles. A supersonic experiment determined drag on a 2.59g projectile from M1.2 to M2.9 using optical chronographs to measure initial and final velocities over a separation of 91.44m. This experiment supported direct proportionality within the experimental error of 1%-2%.

## Acknowledgements

The authors are grateful to Louisiana Shooters Unlimited for use of their range facilities in Lake Charles, Louisiana, and Dragonman's Range for use of their shooting range in Colorado Springs, Colorado. This work was funded by the US Air Force Academy and BTG Research who played no role in manuscript preparation other than employment of the authors.

**General Nomenclature**

$C_d$  drag coefficient
$A$  cross-sectional area
$V$  flow velocity
$\rho$  air density
$F_d$  drag force
$m$  projectile mass
$d$  distance between chronographs
$v_f$  final projectile velocity as measured by far chronograph
$v_i$  initial projectile velocity as measured by near chronograph
$v$  average projectile velocity over the interval